\documentclass[trackchanges,twocolumn,floatfix]{aastex701}
\usepackage{graphicx}	
\usepackage{amsmath}	
\usepackage{hyperref}
\usepackage{cprotect}
\usepackage{multirow}
\usepackage{placeins} 
\usepackage{orcidlink}
\usepackage{wasysym}
\usepackage{float}
\usepackage{tabularx}
\usepackage{booktabs}

\usepackage[version=4]{mhchem}
\usepackage{wrapfig}

\shorttitle{Supercritical EOS}
\shortauthors{Marcum et al. }

\newcounter{rxn}

\begin{document}

\title{An Equation of State for Supercritical Silicate-Hydrogen Mixtures at Sub-Neptune Interior Conditions}

\correspondingauthor{Sarah P. Marcum}

\author[orcid=0009-0001-6858-9849,sname='Marcum']{Sarah P. Marcum}
\affiliation{Department of Earth, Planetary, and Space Sciences, University of California, Los Angeles, CA 90095, USA}
\email[show]{smarcum13@g.ucla.edu}

\author[orcid=0000-0003-3778-2432,sname='Stixrude']{Lars Stixrude}
\affiliation{Department of Earth, Planetary, and Space Sciences, University of California, Los Angeles, CA 90095, USA}
\email{lstixrude@epss.ucla.edu}

\author[orcid=0000-0002-1299-0801,sname='Young']{Edward D. Young}
\affiliation{Department of Earth, Planetary, and Space Sciences, University of California, Los Angeles, CA 90095, USA}
\email{eyoung@epss.ucla.edu}

\begin{abstract}
Many sub-Neptunes are expected to contain long-lived molten silicate
interiors beneath dense H$_2$-rich envelopes. At pressures and temperatures
near the atmosphere--interior boundary, MgSiO$_3$ and H$_2$ may become fully
miscible, forming a supercritical silicate--hydrogen fluid. Here we use
density functional theory molecular dynamics simulations to construct a
self-consistent equation of state for supercritical MgSiO$_3$H$_4$,
corresponding to 3.86 wt\% hydrogen expressed as equivalent H$_2$. We fit
the simulation results with a Helmholtz free-energy formulation that yields
density, entropy, heat capacity, thermal expansivity, bulk modulus, and
Gr\"uneisen parameter from a single thermodynamic surface. We find that the
hydrogen-bearing fluid is lower in density than dry MgSiO$_3$ liquid and
deviates significantly from ideal specific-volume mixing between MgSiO$_3$
and H$_2$. The resulting excess volume varies with pressure, indicating
that the interaction between hydrogen and the silicate framework evolves
with compression. Structural analysis shows increasing Si--H coordination
and decreasing persistent H--H bonding at high pressure, consistent with a
transition away from molecular H$_2$-like bonding toward a more strongly
coupled silicate--hydrogen fluid. We incorporate the MgSiO$_3$H$_4$ equation
of state into a composition-dependent MgSiO$_3$--H lookup table and apply it
to representative sub-Neptune structure models. The models show that
hydrogen partitioning between the atmosphere and condensed interior affects
the planetary adiabat relative to the MgSiO$_3$ liquidus. These results demonstrate that
supercritical silicate--hydrogen fluids have distinct thermodynamic and
structural properties that must be accounted for when modeling
sub-Neptune interiors.
\end{abstract}


\section{Introduction}

Sub-Neptunes are among the most abundant classes of exoplanets identified in
the current census of the Galaxy \citep[e.g.,][]{fressin2013a,Fulton2017a}.
These planets have radii intermediate between Earth and Neptune and bulk
densities of 1--3~g~cm$^{-3}$, lower than expected for compressed
rock and metal alone. Interpreting their interiors therefore depends
fundamentally on the equations of state assigned to candidate planetary
materials. Equations of state for silicates, iron metal, water, and H$_2$
have been used to infer the compositions and layer mass fractions of
sub-Neptunes from measured masses and radii
\citep[e.g.,][]{Seager2007,Valencia2006,Sotin2007,RogersSeager2010,
LopezFortney2014,Zeng2016,NixonMadhusudhan2021,Aguichine2021,Vazan2022}.
Because similar masses and radii can be produced by different combinations
of silicate, metal, water, and H$_2$, the inferred interior structure depends
strongly on the adopted equations of state.

The low densities of sub-Neptunes have commonly been interpreted either as
evidence for volatile-rich, water-bearing interiors
\citep[e.g.,][]{Fortney2007,Zeng2019}, or as rocky planets with molten or
partially molten interiors overlain by H$_2$-rich primary atmospheres
\citep{Chachan2018,Bean2021}. In the latter picture, even percent-level
H$_2$ envelopes can strongly influence the observed radii and bulk
densities of sub-Neptunes. Such envelopes also affect the thermal evolution
of the planet: dense H$_2$-rich atmospheres can slow cooling and allow
initially molten silicate interiors to persist over Gyr timescales
\citep{Ginzburg2016a,Misener2022,Rogers2024}. As a result, many
observed sub-Neptunes may contain long-lived magma oceans beneath their
envelopes.

If magma oceans persist beneath H$_2$-rich atmospheres, chemical exchange
between the atmosphere and condensed interior is expected. Hydrogen can
dissolve into molten silicate, while silicate-derived species may be
transferred into the overlying envelope
\citep[e.g.,][]{Kite2019,Schlichting_Young_2022,Misener2023, Charnoz2023,Young_2024,RogersYoungSchlichting2025_MNRAS, Nixon2025_TOI270d,werlen_sub-neptunes_2025}.

The traditional picture of a sharp boundary between a silicate magma ocean
and an H$_2$-rich atmosphere may be incomplete at sub-Neptune interior
conditions. Recent work by
\citet{gilmore_core-envelope_2025} show that MgSiO$_3$ and H$_2$ can become fully miscible at the relevant pressures and temperatures. In this case, the
atmosphere--interior boundary corresponds to a phase change from a supercritical magma ocean composed of silicate and hydrogen and a separate hydrogen-rich phase 
\citep{Markham2022,Young_2024}. Interior to this boundary, the condensed
region is a hydrogen-bearing supercritical fluid whose thermodynamic properties must be determined directly. Recent population models further emphasize that hydrogen--silicate--metal miscibility may influence the structures and demographics of sub-Neptunes by controlling whether planets retain fully miscible interiors \citep{young_influences_2026}.

This miscible regime introduces a material property problem for planetary
structure models. Existing treatments commonly rely on ideal volume
mixing and endmember equations of state for dry silicate, H$_2$, and other
components. Such approaches are unlikely to accurately capture the density of silicate-hydrogen fluids. 
Moreover, thermodynamic derivatives such as heat capacity, thermal expansivity, bulk
modulus, and the Gr\"uneisen parameter control the interior adiabat and
therefore influence the density and thermal structure inferred from
mass--radius models. A self-consistent equation of state for supercritical
MgSiO$_3$--H mixtures is therefore needed to connect the phase-equilibrium
picture of sub-Neptune interiors to planetary structure calculations.

In this work, we use first-principles molecular dynamics simulations to construct an
equation of state for supercritical MgSiO$_3$H$_4$ (corresponding to
3.86 wt\% hydrogen expressed as equivalent H$_2$). We fit the simulation
results with a Helmholtz free-energy formulation that yields pressure,
internal energy, entropy, heat capacity, thermal expansivity, bulk modulus,
and Gr\"uneisen parameter from a single thermodynamic surface. To apply the
equation of state to planets with different hydrogen abundances, we combine the
MgSiO$_3$H$_4$ free-energy surface with a dry MgSiO$_3$ liquid equation of state to construct a MgSiO$_3$--H lookup table spanning the compositional interval from dry MgSiO$_3$ to MgSiO$_3$H$_4$. We then compare the resulting material properties with dry MgSiO$_3$ liquid, analyze the local structure of the hydrogen-bearing fluid, and implement our equation of state into a sub-Neptune structure model.

\section{Supercritical Silicate--Hydrogen Equation of State}
\label{EOS}

\subsection{First-Principles Simulations}

We performed ab initio molecular dynamics simulations of supercritical
MgSiO$_3$H$_4$ using density functional theory (DFT) in the
Born--Oppenheimer approximation. The
simulations were carried out with the projector augmented wave (PAW) method
as implemented in the Vienna Ab initio Simulation Package (VASP)
\citep{Kresse1996}, using the PBEsol exchange--correlation functional
\citep{Perdew2008}. Each simulation cell contains 243 atoms: 27 Mg, 27 Si,
81 O, and 108 H atoms. This composition corresponds to MgSiO$_3$H$_4$, or
3.86 wt\% hydrogen by mass. 

\begin{figure}[h]
\centering
   \includegraphics[width=0.45\textwidth]{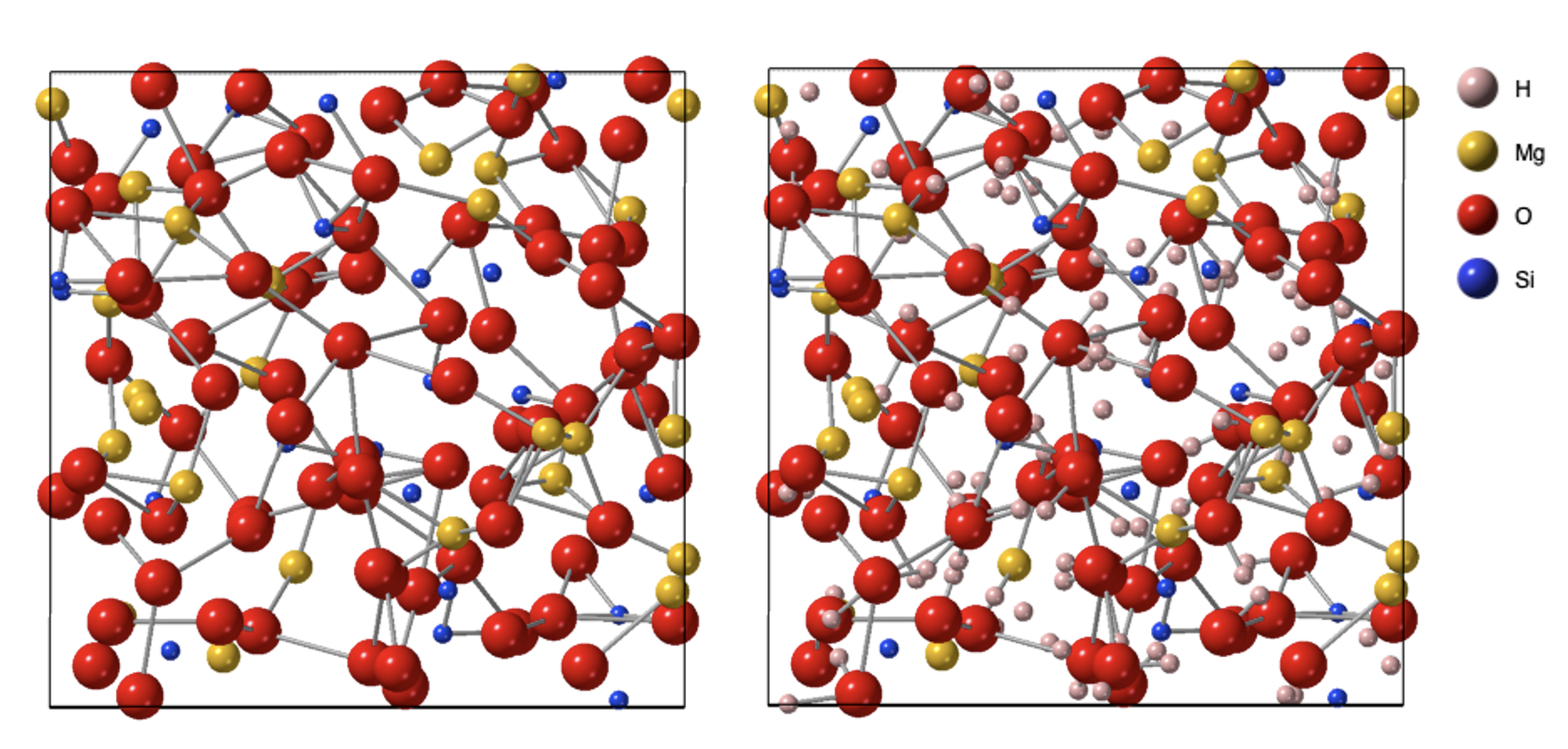}
    \caption{Initial simulation configuration before and after hydrogen
    insertion. Hydrogen was added to a pre-equilibrated MgSiO$_3$ liquid
    structure, producing the MgSiO$_3$H$_4$ composition simulated in this
    work. This composition corresponds to 3.86 wt\% hydrogen by mass.}
\label{fig:simulation}
\end{figure}

Initial liquid structures were generated by melting crystalline MgSiO$_3$
in a cubic cell for 10 ps at 8000 K and approximately 4 GPa. Hydrogen was
then added to the liquid configuration by randomly inserting 108 H atoms
into available voids of the simulation cell, with a minimum separation of
1.5~\AA{} from all other atoms to avoid nonphysical overlaps
(Figure~\ref{fig:simulation}).

The hydrogen-bearing structures were equilibrated for an additional 10 ps at
the target isotherm temperature, 4000, 6000, or 8000 K, starting from volumes
corresponding to pressures of a few GPa. We used a timestep of 0.5 fs,
consistent with recent first-principles simulations of hydrogen-bearing
systems \citep{Gupta2024_water,gilmore_core-envelope_2025}. Isothermal
compression paths were then constructed by reducing the cell volume stepwise
and running simulations for 10--15 ps at each volume.

\begin{figure*}[t]
\centering
    \includegraphics[width=0.95\textwidth]{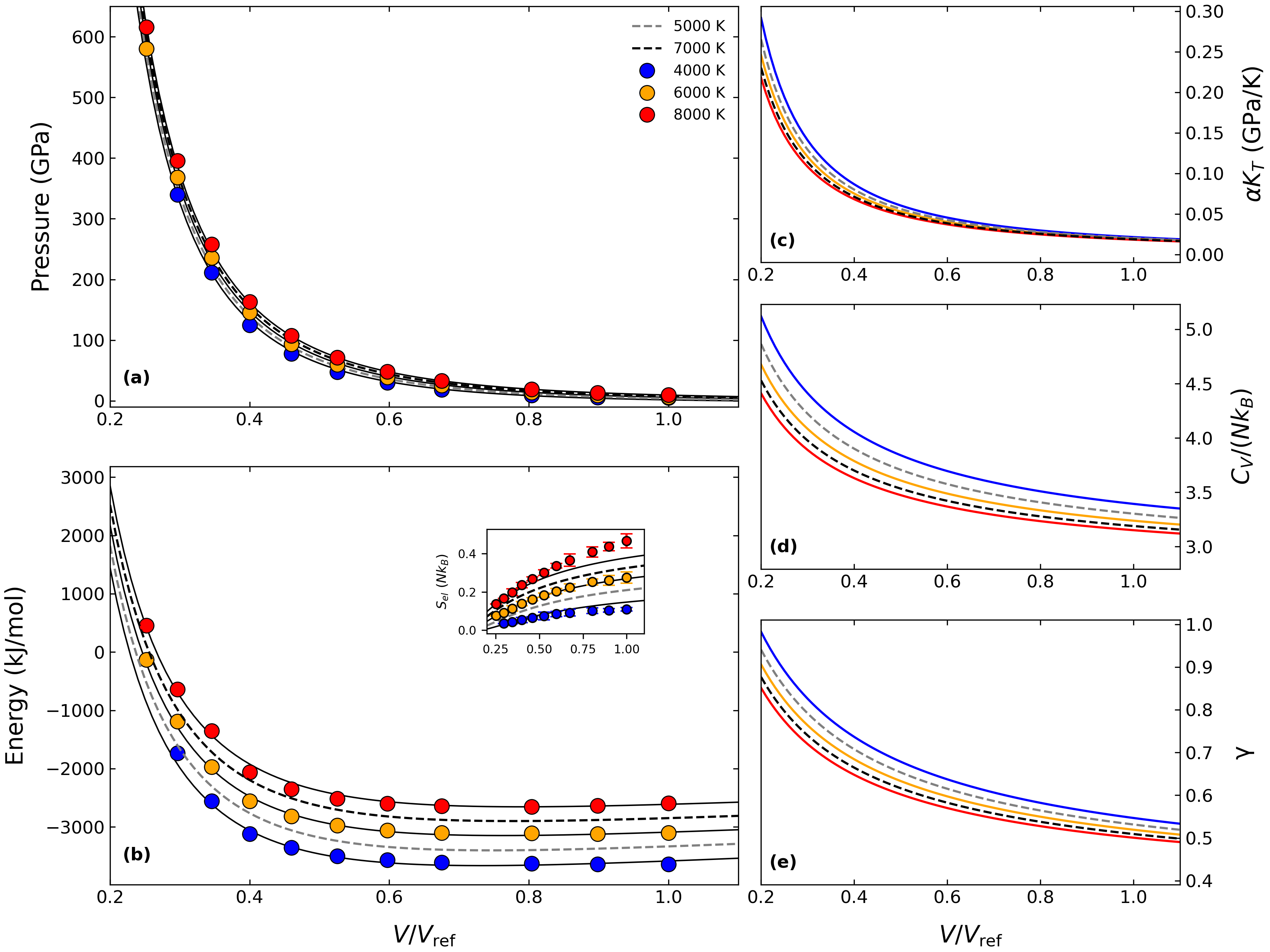}
    \caption{Our simulations results and our global EOS fit for supercritical
MgSiO$_3$H$_4$. Panels (a) and (b) show pressure and internal energy along
the 4000, 6000, and 8000 K isotherms; solid lines are the EOS fit and
filled circles are DFT-MD simulation results. The inset in panel (b) shows
the electronic entropy, $S_{\rm el}$. Panels (c)--(e) show
derived thermodynamic properties calculated from the same fitted
free-energy surface: the thermal pressure coefficient, $\alpha K_T$,
isochoric heat capacity, $C_V$, and Gr\"uneisen parameter, $\gamma$.
Dashed gray lines show interpolation to 5000 and 7000 K.}
\label{fig:eos_summary}
\end{figure*}

All simulations were performed in the canonical (NVT) ensemble using a
Nos\'e--Hoover thermostat \citep{Nose1984,Hoover1985}. The Brillouin zone was
sampled at the $\Gamma$ point and a plane-wave cutoff energy of 500 eV was
used. These settings yield total energy and pressure convergence within
10 meV atom$^{-1}$ and 0.5 GPa, respectively. Electronic thermal effects were
included using the Mermin functional \citep{Mermin1965}, with the electronic
temperature set equal to the ionic temperature. Thermodynamic averages were
calculated after discarding the first 20\% of each trajectory, and
uncertainties were estimated using the blocking method \citep{Flyvbjerg1989}.

\subsection{Supercritical Equation of State}

To describe the thermodynamic properties of the simulated supercritical
MgSiO$_3$H$_4$ fluid, we adopt the liquid equation-of-state formalism of
\citet{dekoker2009}. In this approach, pressure, internal energy, entropy, heat capacity, and other thermodynamic properties are derived from a single fitted Helmholtz free-energy surface,
\begin{equation}
    F(V,T) = F_{\mathrm{ig}}(V,T) + F_{\mathrm{xs}}(V,T) + F_{\mathrm{el}}(V,T).
\end{equation}
where $F_{\rm ig}$ is the ideal atomic contribution, $F_{\rm el}$ is the
thermal electronic contribution, and $F_{\rm xs}$ is the excess
interatomic contribution.

Figure~\ref{fig:eos_summary} shows the fitted EOS compared with the DFT-MD
simulation data along the 4000, 6000, and 8000 K isotherms. The model
reproduces the simulated pressures and internal energies over the sampled pressure range, extending to approximately 600 GPa.

The panels to the right of Figure~\ref{fig:eos_summary} show thermodynamic
derivatives calculated directly from the fitted Helmholtz free-energy
surface. The thermal pressure coefficient, $\alpha K_T$, isochoric heat
capacity, $C_V$, and Gr\"uneisen parameter, $\gamma$, vary smoothly across the sampled pressure--temperature range. This regular behavior is important for planetary structure calculations, where the EOS is used to construct isentropic temperature profiles and evaluate density and thermodynamic properties along the condensed-interior adiabat. 

Additional details of the Helmholtz free-energy formulation and the
derivation of thermodynamic quantities are provided in
Appendix A.

\section{Material Properties}
\label{sec:material_properties}

\subsection{Comparison with Dry MgSiO$_3$ Liquid}

We first compare the simulated MgSiO$_3$H$_4$ densities with an ideal
specific-volume mixture of dry MgSiO$_3$ liquid and pure H$_2$ at the same
pressure and temperature. This comparison provides a useful reference
because it represents the density that would be predicted without explicitly
simulating a miscible MgSiO$_3$--H fluid. The dry MgSiO$_3$ and H$_2$
reference densities are taken from \citet{dekoker2009} and
\citet{Chabrier2019}, respectively.

The ideal mixture density is calculated by additive specific-volume mixing
at fixed pressure and temperature; that is, the volume per unit mass of the
mixture is taken to be the mass-weighted sum of the dry MgSiO$_3$ and
H$_2$ specific volumes:
\begin{equation}
\frac{1}{\rho_{\rm ideal}}
=
\frac{1-w_{\rm H}}{\rho_{\rm MgSiO_3}}
+
\frac{w_{\rm H}}{\rho_{\rm H_2}} ,
\end{equation}
where $w_{\rm H}$ is the hydrogen mass fraction of MgSiO$_3$H$_4$,
expressed as an equivalent H$_2$ component.

\begin{figure}
\centering
   \includegraphics[width=0.45\textwidth]{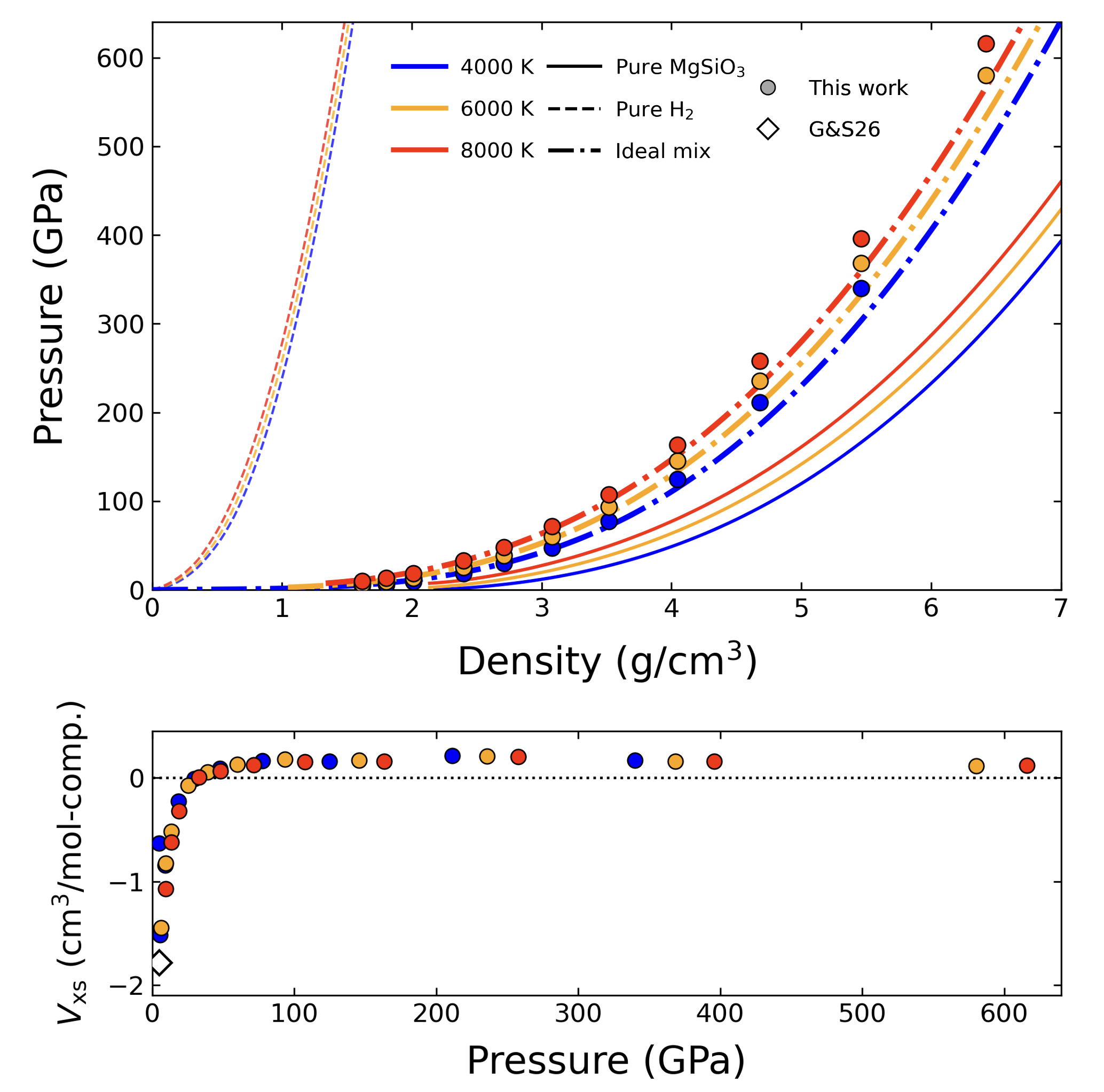}
    \caption{Equation of state and excess-volume comparison for MgSiO$_3$H$_4$.
(top) Density of MgSiO$_3$H$_4$ from our simulations (circles) compared with
ideal mixing between dry MgSiO$_3$ and H$_2$ (dashed lines). Also shown are
the equations of state of dry MgSiO$_3$ \citep{dekoker2009} and H$_2$
\citep{Chabrier2019}. Colors indicate temperature: 4000 K (blue), 6000 K
(orange), and 8000 K (red). 
(bottom) Excess volume of MgSiO$_3$H$_4$ computed from our simulations
(circles; same color scheme as the top panel), compared with the excess volume
at MgSiO$_3$H$_4$ composition from \citet{gilmore_core-envelope_2025}
(white diamond).}
\label{fig:mix}
\end{figure}

The simulated MgSiO$_3$H$_4$ densities deviate systematically from the
ideal-mixture prediction. The discrepancy reaches several percent at high
pressure and increases with compression, indicating that the density of the
hydrogen-bearing fluid cannot be described as a simple physical mixture of
dry silicate liquid and hydrogen.

The excess volume provides a direct measure of this non-ideality. It is
defined as the difference between the simulated MgSiO$_3$H$_4$ volume and
the volume predicted by ideal specific-volume mixing. As shown in the lower panel of
Figure~\ref{fig:mix}, the excess volume changes across the simulated
pressure range, demonstrating that the interaction between hydrogen and the
silicate network evolves with compression. This behavior is consistent with
the results of \citet{gilmore_core-envelope_2025}, who also find non-ideal
volumes of solution in miscible silicate--hydrogen fluids. The pressure
dependence of the excess volume provides a thermodynamic measure of the
same structural changes discussed below, including the breakdown of H--H
bonding and the increasing interaction between hydrogen and the silicate
framework.

\subsection{Radial Distribution Function}

We use radial distribution functions (RDFs) to characterize the local
structure of the simulated supercritical MgSiO$_3$H$_4$ fluid. Following
standard definitions \citep{McQuarrie1976_StatisticalMechanics}, for each atomic pair,
$A$--$B$, we compute
\begin{equation}
g_{AB}(r) =
\frac{V}{4\pi r^2 N_A N_B}
\left\langle
\sum_{i \in A}
\sum_{\substack{j \in B \\ j \neq i}}
\delta(r-r_{ij})
\right\rangle,
\end{equation}
where $V$ is the simulation-cell volume, $N_A$ and $N_B$ are the numbers of
atoms of species $A$ and $B$, $r_{ij}$ is the distance between atoms $i$ and $j$, and the angle
brackets indicate time averaging. RDFs are averaged over the equilibrated portion of each
trajectory.

\begin{figure}[h]
\centering
   \includegraphics[width=0.45\textwidth]{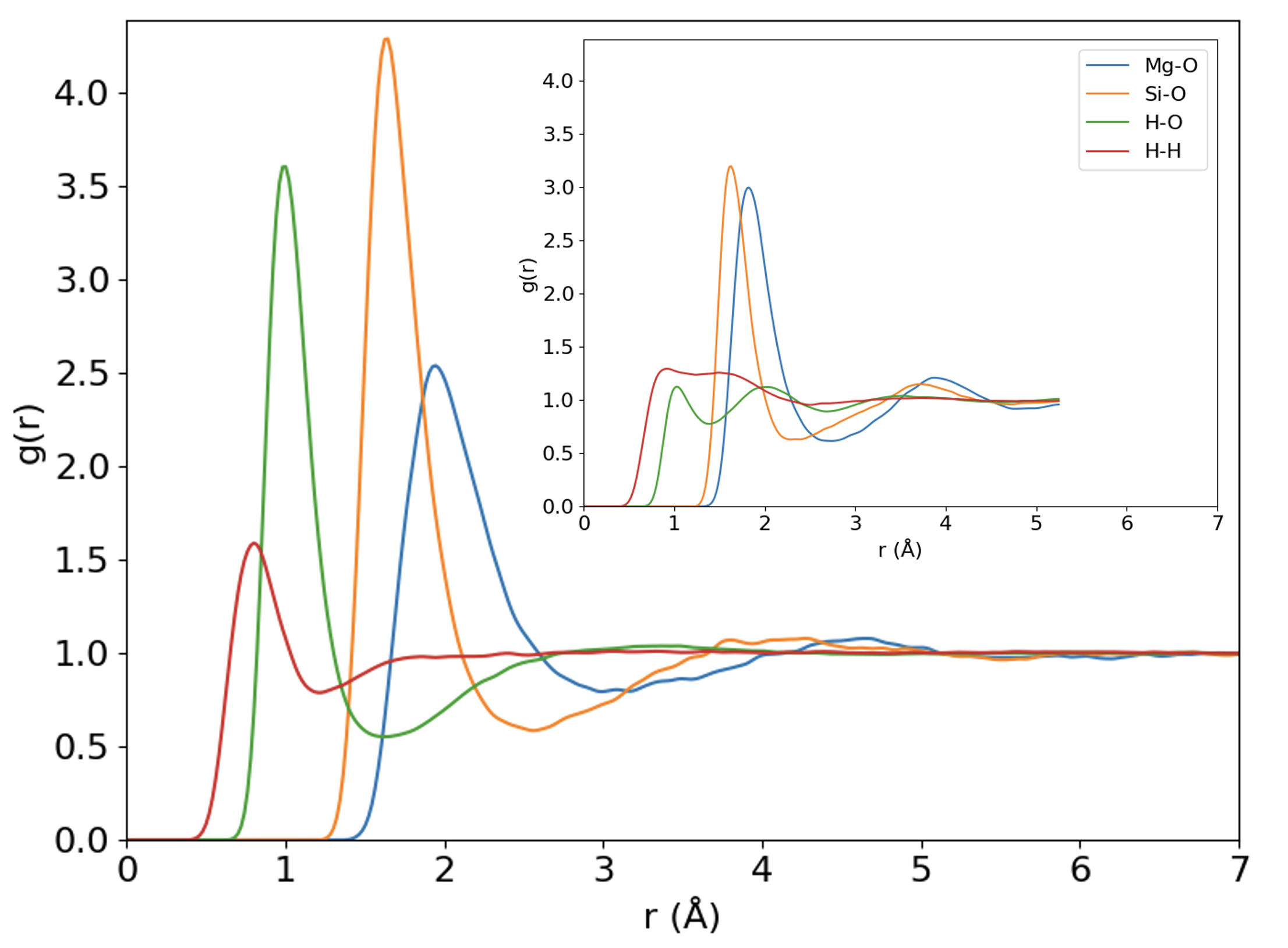}
    \caption{Radial distribution functions for Si--O, Mg--O, O--H, and H--H
pairs at 6000~K. The main panel shows the low-pressure simulation at
6.40~GPa, and the inset shows the high-pressure simulation at
580.32~GPa.}
\label{fig:rdf}
\end{figure}

Figure~\ref{fig:rdf} shows representative RDFs along the 6000~K isotherm at 6.4 GPa and 580 GPa.
The Si--O and Mg--O correlations retain distinct first-neighbor peaks over
the pressure range considered, indicating persistent short-range silicate
structure in the supercritical fluid. Hydrogen also participates in the
short-range structure, as shown by O--H and H--H correlations. The most
pronounced pressure-induced change is in the H--H RDF: the sharp
low-pressure H--H first-neighbor peak associated with molecular hydrogen is
strongly reduced at high pressure. This indicates breakdown of molecular
H--H bonding and a transition from a more molecular hydrogen environment
toward one with more atomic hydrogen within the silicate fluid.

\subsection{Coordination and Hydrogen Persistence}

From the RDFs, we determine coordination numbers by integrating the first
peak:
\begin{equation}
{\rm CN}_{A-B} =
4\pi \rho_B
\int_0^{r_{\rm cut}}
g_{AB}(r) r^2 \, dr,
\end{equation}
where $\rho_B$ is the number density of species $B$, and $r_{\rm cut}$ is
the location of the first minimum after the primary peak in $g_{AB}(r)$.

\begin{figure}
\centering
\includegraphics[width=0.45\textwidth]{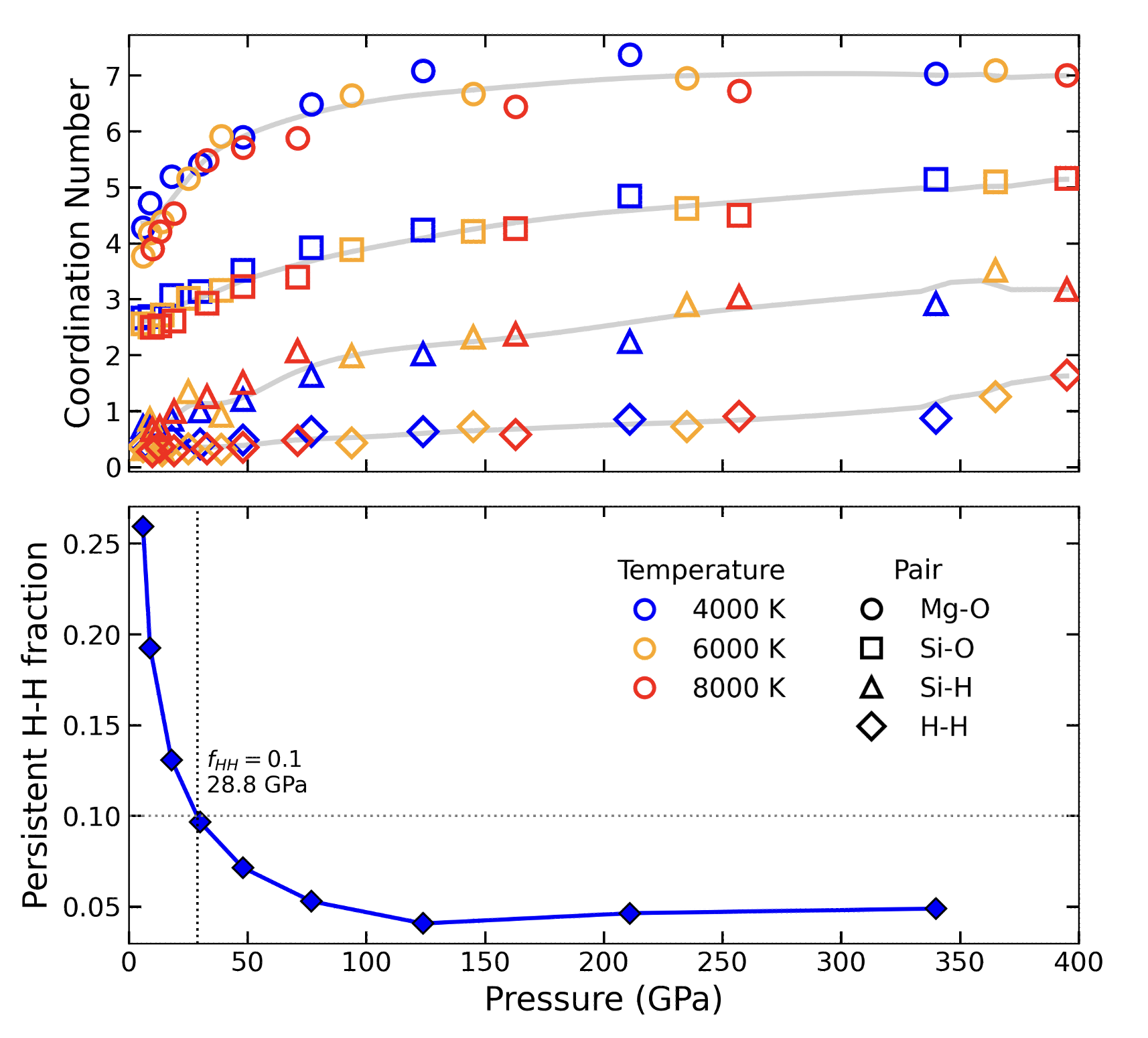}
\caption{Coordination numbers and H--H persistence in supercritical
MgSiO$_3$H$_4$. Top: Mg--O, Si--O, Si--H, and H--H coordination numbers
as a function of pressure. Symbols show individual simulations colored by
temperature; gray curves show temperature-averaged trends. Bottom:
time-averaged fraction of H atoms in persistent H--H contacts at
4000~K. Persistent contacts are defined as H--H pairs that remain shorter
than the first minimum in the H--H RDF for at least 100~fs.}
\label{fig:cn}
\end{figure}

Figure~\ref{fig:cn} shows the Mg--O, Si--O, Si--H, and H--H coordination
numbers as a function of pressure, together with the fraction of hydrogen
atoms participating in persistent H--H contacts at 4000~K. For the
RDF-derived coordination numbers, coordination generally increases with
compression, as expected from the increasing density of the fluid. However,
Si--O coordination remains below 5 over the pressure range explored here.
This behavior differs from dry MgSiO$_3$ and related silicate liquids, in
which Si--O coordination increases more rapidly with pressure and is known
to exceed 6 under comparable high-pressure conditions
\citep{StixrudeKarki2005_Science,Sun2011_Diopside,Sanloup2013_MoltenBasalt,Luo_Deng_2025}.

The limited pressure-induced growth in Si--O coordination in the
H-bearing fluid, compared with pure MgSiO$_3$ liquid, indicates that
hydrogen changes how the silicate framework responds to compression. The
origin of this effect is increasing Si--H bonding. Si--H bonds are present
at all pressures, and the Si--H coordination number increases with
pressure. Our results are consistent with those of
\citet{gilmore_core-envelope_2025}, who also find that miscible
silicate--hydrogen fluids contain a substantial population of Si--H bonds.

We perform a pair-persistence analysis to distinguish long-lived H--H
contacts from transient near-neighbor encounters. For each simulation, an
H--H pair is first identified using the same RDF-based criterion used for
the coordination analysis: the H--H distance must lie within the first
minimum in the H--H RDF. Along the 4000~K isotherm, this cutoff decreases
from $r_{\rm HH,cut}=1.29$~\AA{} at 6~GPa to
$r_{\rm HH,cut}=1.07$~\AA{} at 340~GPa. An H--H contact is then considered
persistent only if the same pair remains continuously within this cutoff
for at least 100~fs. We report the time-averaged fraction of H atoms
participating in at least one such persistent H--H contact.

The H--H coordination number increases with pressure, but the persistence
analysis shows that this trend does not correspond to increasing molecular
H$_2$ abundance. At 4000~K, the persistent H--H fraction decreases rapidly
with pressure and falls below 0.1 near the same pressure range as the
$V^{\rm xs}=0$ crossover in Figure~\ref{fig:mix}. Thus, high-pressure
H--H coordination primarily reflects compression-driven near-neighbor
proximity rather than long-lived molecular H$_2$.

\section{Supercritical Interior Models}
\label{Planet}

\subsection{MgSiO$_3$--H Lookup Table}
\label{sec:lookup_table_eos}

To incorporate hydrogen-bearing silicate fluids into planetary structure
calculations, we constructed a composition-dependent MgSiO$_3$--H lookup EOS
using HeFESTo \citep{hefesto2011}. The lookup table is based on two liquid
Helmholtz free-energy surfaces: the dry MgSiO$_3$ liquid EOS of
\citet{dekoker2009} and the supercritical MgSiO$_3$H$_4$ EOS fit in this
work. These two compositions are treated as endmembers for representing
intermediate MgSiO$_3$--H compositions.

Each endmember is described by a Helmholtz free energy, $F(V,T)$. At a
specified pressure and temperature, HeFESTo evaluates the corresponding
Gibbs free energy by finding the volume that satisfies the target pressure
and applying the Legendre transform
\begin{equation}
    G(P,T) = F(V,T) + PV .
\end{equation}
The composition-dependent fluid EOS is then constructed in Gibbs
free-energy space, which is the natural thermodynamic potential for
fixed-pressure, fixed-temperature calculations.

Intermediate compositions are represented as
\begin{equation}
    \mathrm{MgSiO_3H}_n,
    \qquad 0 \leq n \leq 4,
\end{equation}
where $n$ is the number of hydrogen atoms per MgSiO$_3$ formula unit. Thus
$n=0$ corresponds to dry MgSiO$_3$, while $n=4$ corresponds to the
MgSiO$_3$H$_4$ endmember simulated in this work. Equivalently, we define
the MgSiO$_3$H$_4$ endmember fraction as
\begin{equation}
    X = \frac{n}{4},
    \qquad 0 \leq X \leq 1 .
\end{equation}

The hydrogen mass fraction corresponding to a given stoichiometric
composition is
\begin{equation}
    w_{\rm H}
    =
    \frac{nM_{\rm H}}
    {M_{\rm MgSiO_3}+nM_{\rm H}},
    \label{eq:wH_n}
\end{equation}
where $M_{\rm H}$ is the molar mass of atomic hydrogen
($1.008$~g~mol$^{-1}$) and $M_{\rm MgSiO_3}$ is the molar mass of one
MgSiO$_3$ formula unit ($100.39$~g~mol$^{-1}$). Written
in terms of the MgSiO$_3$H$_4$ endmember fraction, this is equivalently
\begin{equation}
    w_{\rm H}
    =
    \frac{X(4M_{\rm H})}
    {(1-X)M_{\rm MgSiO_3}
    +
    X(M_{\rm MgSiO_3}+4M_{\rm H})}.
    \label{eq:wH_X}
\end{equation}
The MgSiO$_3$H$_4$ endmember therefore corresponds to approximately
3.86 wt\% H.

For the lookup table, the MgSiO$_3$--H fluid is treated as an ideal solution
between dry MgSiO$_3$ and MgSiO$_3$H$_4$. In terms of the endmember
fraction $X$, the mixed liquid Gibbs free energy is
\begin{equation}
\begin{split}
    G_{\rm liq}
    ={}&
    (1-X)G_{\rm dry}
    +
    XG_{\rm hyd} \\
    &+
    RT
    \left[
    (1-X)\ln(1-X)
    +
    X\ln X
    \right],
    \label{eq:Gliq_ideal_simple}
\end{split}
\end{equation}
where $G_{\rm dry}$ and $G_{\rm hyd}$ are the Gibbs free energies of the dry
MgSiO$_3$ and MgSiO$_3$H$_4$ endmembers, respectively. Density and the other
thermodynamic properties are then derived from the mixed Gibbs free-energy
surface.

The final EOS table is indexed by hydrogen concentration, pressure, and
temperature. For each point in $(w_{\rm H},P,T)$ space, the table reports
density, entropy, enthalpy, heat capacity, thermal expansivity, adiabatic
bulk modulus, reference density, compression ratio, and Gr\"uneisen
parameter. The table spans the binary interval from dry MgSiO$_3$ to
MgSiO$_3$H$_4$, corresponding to approximately 0--3.86 wt\% H.

The MgSiO$_3$--H equation of state table is
available at \url{https://github.com/s-marcum/MgSiO3-H-EOS}.

\subsection{Planet Model Setup}

We use the PlanetLab model as a vehicle for applying the MgSiO$_3$--H
lookup EOS to sub-Neptune interior structure calculations
\citep{Young2025_Differentiation}. The details of the model have been
described previously, but we summarize the aspects most relevant to the
condensed interior here. The planet is modeled as a spherically symmetric
body consisting of a well-mixed condensed interior overlain by an
H$_2$-rich atmosphere. The condensed interior is assumed to be single phase
over the pressure--temperature range considered. Its hydrogen abundance is
set by mass balance between the total hydrogen inventory and the hydrogen
contained in the overlying atmosphere.

For a specified total H$_2$ inventory and atmosphere--interior boundary
pressure, the atmospheric mass fraction is determined self-consistently
rather than prescribed independently. The atmosphere is integrated outward
from the MgSiO$_3$--H$_2$ binodal, which sets the pressure, temperature,
and gas composition at the base of the envelope. In hydrostatic form, the
boundary pressure corresponds to the weight of the overlying atmosphere,
\begin{equation}
P_{\rm surf}
=
\int_{R_{\rm surf}}^{\infty}
\rho_{\rm atm}(r,T,x_i)\,g(r)\,dr ,
\label{eq:atm_weight}
\end{equation}
where $R_{\rm surf}$ is the radius of the condensed interior, $R_{\rm out}$
is the outer atmosphere, and $x_i$ denotes the gas composition. The
envelope is H$_2$-rich rather than strictly pure H$_2$: near the
atmosphere--interior boundary, its composition is set by equilibrium with
the condensed phase, while the outer atmosphere is H$_2$ dominated. The
resulting atmospheric hydrogen mass is subtracted from the total hydrogen
inventory, and the remaining hydrogen sets the fixed composition of the
condensed MgSiO$_3$--H interior.

The condensed structure is calculated by solving the standard equations of
mass conservation and hydrostatic equilibrium \citep[e.g.,][]{Seager2007},
\begin{equation}
\frac{dm}{dr} = 4\pi r^2\rho,
\label{eq:dmdr}
\end{equation}
\begin{equation}
\frac{dP}{dr} = -\frac{Gm\rho}{r^2},
\label{eq:dPdr}
\end{equation}
where $m$ is the mass enclosed within radius $r$, $\rho$ is density, $P$ is
pressure, and $G$ is the gravitational constant. For the condensed
interior, the density is obtained using the MgSiO$_3$--H equation of state,
\begin{equation}
\rho = \rho_{\rm EOS}(w_{\rm H},P,T),
\label{eq:rho_eos}
\end{equation}
where $w_{\rm H}$ is the hydrogen mass fraction in the condensed
MgSiO$_3$--H phase. This hydrogen abundance is held fixed throughout the
condensed interior for a given model, consistent with efficient convection
and complete miscibility in the supercritical regime.

Because the condensed interior is fluid and expected to convect efficiently,
its thermal profile is assumed to be isentropic. In the constant-entropy
implementation, the adiabat is anchored at the atmosphere--interior boundary
condition, $(P_{\rm surf},T_{\rm surf})$. The entropy at this boundary
defines the entropy of the condensed interior,
\begin{equation}
S_{\rm surf} =
S_{\rm EOS}(w_{\rm H},P_{\rm surf},T_{\rm surf}).
\label{eq:Ssurf}
\end{equation}
At each deeper pressure, the adiabatic temperature is then found by solving
\begin{equation}
S_{\rm EOS}(w_{\rm H},P,T_{\rm ad}) = S_{\rm surf}.
\label{eq:entropy_adiabat}
\end{equation}
The resulting function $T_{\rm ad}(P)$ is used during the interior structure
calculation. For each trial central pressure, the structure equations are
integrated through the condensed mass while density is evaluated from the
EOS along the precomputed constant-entropy path. The central pressure is
adjusted until the integrated outer boundary pressure matches the
atmosphere-imposed surface pressure. This procedure yields self-consistent
radial profiles of pressure, temperature, density, and thermodynamic
properties for the condensed interior. 

\subsection{Planet Models}

In Figure 6, we compare two representative sub-Neptune models with the same planet mass,
$M_p=6.0\,M_{\oplus}$, the same equilibrium temperature,
$T_{\rm eq}=1000$~K, and the same total H$_2$ inventory of 4.0 wt\%.
The models differ only in the pressure of the atmosphere--interior boundary,
which is set by the MgSiO$_3$--H$_2$ binodal \citep{gilmore_core-envelope_2025}. This boundary pressure
controls how the fixed hydrogen inventory is partitioned between the
H$_2$-rich atmosphere and the supercritical MgSiO$_3$--H condensed interior.

In model A, the atmosphere--interior boundary occurs at
$P_{\rm surf}=1$~GPa and $T_{\rm surf}=3746$~K. The atmosphere contains
0.60\% of the planet mass, while the condensed interior contains
3.44 wt\% H$_2$ equivalent. In model B, the boundary
occurs at $P_{\rm surf}=5.0$~GPa and $T_{\rm surf}=3204$~K. The atmosphere
is more massive, comprising 2.50\% of the planet mass, and the condensed
interior contains 1.52 wt\% H$_2$ equivalent.

\begin{figure*}[t]
\centering
    \includegraphics[width=0.95\textwidth]{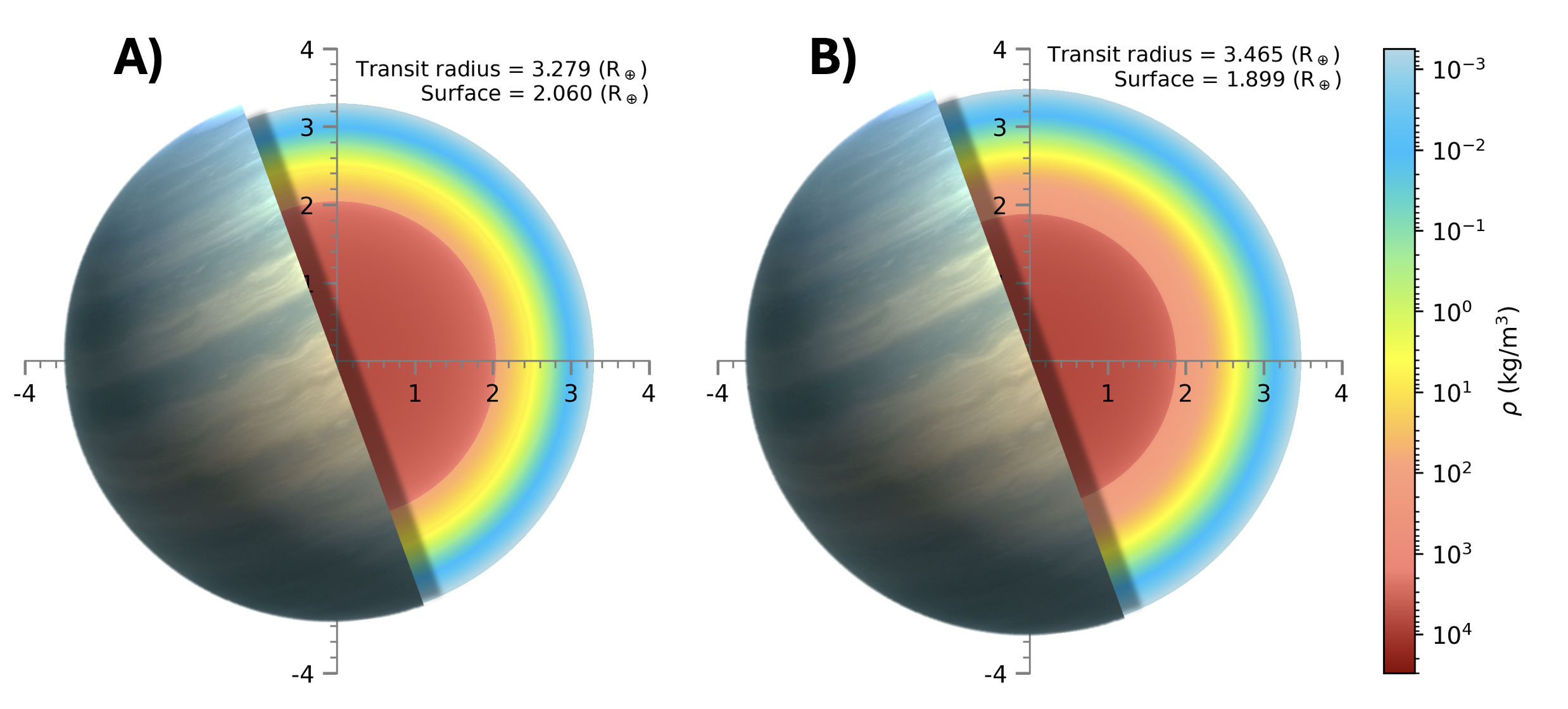}
    \caption{Comparison of two $6.0\,M_{\oplus}$ sub-Neptune models with the
same equilibrium temperature and total H$_2$ mass fractions of 4$\%$, but different
MgSiO$_3$--H$_2$ binodal pressures. The lower-pressure model
($P_{\rm surf}=1$~GPa; model A) retains more hydrogen in the
supercritical MgSiO$_3$--H condensed interior, while the higher-pressure
model ($P_{\rm surf}=5.0$~GPa; model B) partitions more hydrogen into the atmosphere.}
\label{fig:planet_comparison}
\end{figure*}

In planet A, the larger
hydrogen abundance in the condensed phase lowers the interior density and
increases the condensed-interior radius to $2.060\,R_{\oplus}$. In planet B, more hydrogen resides in the atmosphere, leaving a
comparatively drier and denser condensed interior with a smaller radius of
$1.899\,R_{\oplus}$. The bulk density of the condensed interior increases
from 3.76~g~cm$^{-3}$ to 4.71~g~cm$^{-3}$.

The atmospheric structure changes in the opposite direction. Planet A has a smaller H$_2$ atmosphere and a transit radius of
$3.279\,R_{\oplus}$, while planet B has a more massive
atmosphere and a transit radius of $3.465\,R_{\oplus}$. The two transit radii
differ by $0.186\,R_{\oplus}$, or about 5.7\%. This fractional difference
is larger than the few-percent radius uncertainties achieved for many
well-characterized sub-Neptune-size planets
\citep{fulton2018a,vaneylen2018a}. The comparison illustrates that
hydrogen affects sub-Neptune structure through both reservoirs: it can inflate
the observable atmosphere and, when dissolved into the supercritical interior,
change the density and radius of the condensed planet itself.

To assess the melting state of the modeled sub-Neptune interiors, we compare the planetary temperature profiles with the MgSiO$_3$ melting curve from \citet{Fei2021}. The effect of dissolved hydrogen is incorporated through the activity of MgSiO$_3$,
\begin{equation}
    a_{\mathrm{MgSiO_3}}
    =
    \gamma_{\mathrm{MgSiO_3}}
    X_{\mathrm{MgSiO_3}},
\end{equation}
where $X_{\mathrm{MgSiO_3}}$ is the mole fraction of MgSiO$_3$ and $\gamma_{\mathrm{MgSiO_3}}$ is its activity coefficient. The liquidus temperature is calculated as
\begin{equation}
    T_{\mathrm{m,H}}
    =
    \frac{T_{\mathrm{m,0}}}
    {1-R\ln\left(a_{\mathrm{MgSiO_3}}\right)/\Delta S_m},
\end{equation}
where $T_{\mathrm{m,0}}$ is the dry MgSiO$_3$ melting temperature and $\Delta S_m$ is the entropy of melting.  Folllowing previous first principles simulation studies \citet{StixrudeKarki2005,Deng2023_MgSiO3MeltingMLP}, we adopt $\Delta S_m/R=7.5$.

The activity coefficient is obtained from a binary fit to the asymmetric subregular-solution model for MgSiO$_3$--H$_2$ mixing by \citet{gilmore_core-envelope_2025} at conditions near the low-pressure silicate--hydrogen binodal.

\begin{figure}[h]
\centering
   \includegraphics[width=0.45\textwidth]{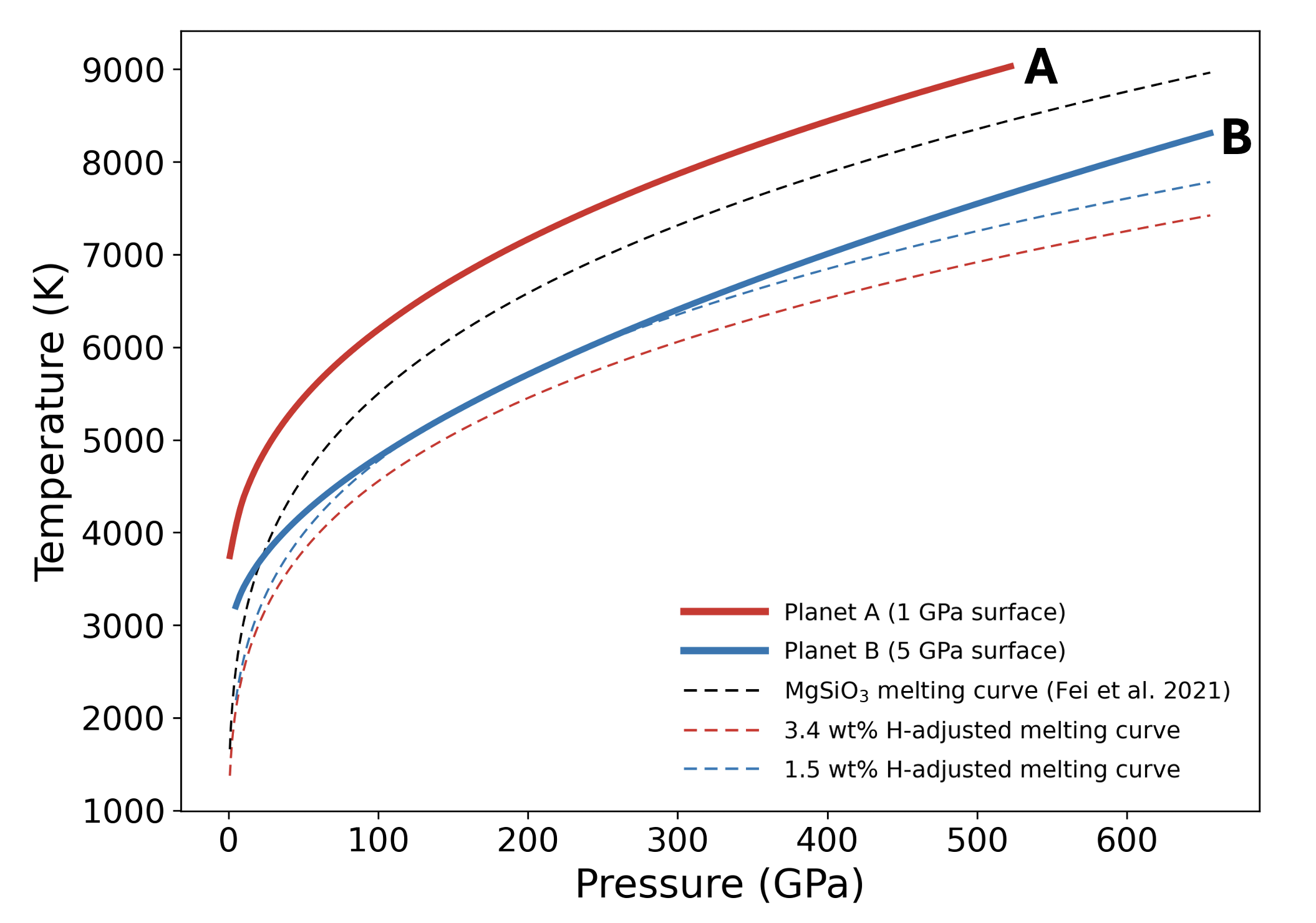}
    \caption{Pressure--temperature profiles of the interior portions for the two modeled sub-Neptunes compared with the dry MgSiO$_3$ melting curve of \citet{Fei2021} and the corresponding hydrogen-adjusted melting curves for representative non-ideal MgSiO$_3$--H$_2$ mixing. Planet A has a more hydrogen-rich interior than Planet B.}
\label{fig:melt}
\end{figure}

Both modeled planets remain super-liquidus and therefore fully molten throughout their interiors (Figure~\ref{fig:melt}). Hydrogen depresses the MgSiO$_3$ liquidus in both cases, although the effect differs substantially between the two models. Planet A retains more hydrogen in its interior and remains well above the hydrogen-adjusted melting curve, whereas the more hydrogen-poor planet B lies much closer to the liquidus over a broad pressure range. This contrast highlights the sensitivity of the melting state to interior hydrogen abundance: as less hydrogen is retained at depth, the melting-point depression weakens and the interior approaches conditions favorable for crystallization. Uncertainties in the high-pressure MgSiO$_3$ liquidus and in the magnitude of the hydrogen-induced melting-point depression limit the significance of the exact temperature offsets, but both models remain fully molten for the representative treatment considered here.

Planet B's proximity to the liquidus may become important during subsequent planetary evolution. In the miscible sub-Neptune models of \citet{RogersYoungSchlichting2025_MNRAS}, the binodal surface contracts inward as the planet cools, moving to higher pressures and reducing the extent of the miscible interior. At the same time, hydrogen is progressively transferred from the interior into the overlying envelope. As the interior becomes more hydrogen-poor, the melting-point depression weakens. Thus, the evolution toward a deeper, higher-pressure binodal and a lower interior hydrogen abundance may bring some sub-Neptunes closer to the liquidus, potentially allowing partial crystallization at later evolutionary stages.

\section{Conclusions}

In this study, we used first-principles molecular dynamics simulations to
construct a self-consistent equation of state for supercritical
MgSiO$_3$H$_4$, a hydrogen-bearing silicate fluid relevant to fully miscible
sub-Neptune interiors. Previous work has shown that the MgSiO$_3$--H$_2$
binodal can define the boundary between an H$_2$-rich envelope and a
supercritical silicate--hydrogen interior, but the thermodynamic properties
of the miscible condensed phase have remained poorly constrained. By fitting
the DFT-MD results with a Helmholtz free-energy formulation, we derive
density and thermodynamic derivatives from a single free-energy surface and
construct a MgSiO$_3$--H lookup table spanning the interval from dry
MgSiO$_3$ to MgSiO$_3$H$_4$.

Relative to dry MgSiO$_3$ liquid, the hydrogen-bearing fluid is lower in
density at all pressure and temperature conditions, by an amount that
deviates significantly from ideal mixing between MgSiO$_3$ and pure H$_2$
fluids. The fitted MgSiO$_3$H$_4$ EOS also differs
from dry MgSiO$_3$ liquid in its compressional and thermal derivatives.
These differences are reflected in the self-consistent planetary adiabats
shown in Figure~\ref{fig:melt}: for the representative sub-Neptune models
considered here, both MgSiO$_3$--H temperature profiles remain
super-liquidus throughout the condensed interior, but the more
hydrogen-poor model lies closer to its hydrogen-adjusted liquidus over much
of the modeled pressure range.

The local structure of the fluid provides a microscopic basis for these
thermodynamic differences. With increasing pressure, hydrogen interacts
more strongly with the silicate framework, as shown by increasing Si--H
coordination. At the same time, persistent H--H contacts decrease with
compression, indicating loss of long-lived molecular H$_2$-like bonding.
Thus, the high-pressure fluid is not simply dry MgSiO$_3$ liquid diluted by
molecular hydrogen. Rather, compression reduces persistent
H--H bonding and increases Si--H coordination, indicating that hydrogen
becomes increasingly incorporated into the local silicate bonding
environment. These results support treating fully miscible
MgSiO$_3$--H fluid as a distinct planetary material.

The PlanetLab calculations illustrate how this material behavior affects
sub-Neptune structure. For the same planet mass and total hydrogen
inventory, changing the atmosphere--interior boundary pressure changes the
partitioning of hydrogen between the atmosphere and condensed interior. The
lower-pressure model retains more hydrogen in the condensed phase, producing
a lower-density and larger condensed interior. The higher-pressure model
places more hydrogen in the atmosphere, producing a denser and smaller
condensed interior but a larger atmospheric contribution to the
radius. Hydrogen therefore influences sub-Neptune radii through both
reservoirs: by inflating the H$_2$-rich atmosphere and by modifying the
density and thermal structure of the supercritical condensed interior.

Further progress will require extending this thermodynamic description to
additional chemical components and testing it experimentally. Iron is
particularly important because Fe--silicate--hydrogen interactions may alter
the density, redox state, phase relations, and transport properties of the
supercritical interior. Recent experiments are beginning to test predictions
for H$_2$--silicate--metal interactions at high pressure and temperature,
including hydrogen dissolution in silicate melts and reactions involving
FeO reduction and water production \citep{Miozzi2025,Horn2025_WetPlanets}.
These measurements provide important constraints on the chemistry of
hydrogen-bearing magma oceans, but direct experimental determinations of
density, compressibility, and thermal properties for supercritical
silicate--hydrogen fluids remain scarce. Coordinated experimental and
first-principles studies will be needed to build predictive equations of
state across the full composition, pressure, and temperature range relevant
to sub-Neptunes.

\section{Acknowledgements} 
 E.D.Y.  and L.S. acknowledge financial support from NASA grant 80NSSC24K0544 (Emerging Worlds program).  


\section*{Appendix A: Thermodynamic EOS Formalism}

The MgSiO$_3$H$_4$ EOS was fit using the Helmholtz free-energy framework of
\citet{dekoker2009} which we briefly review here.. The total Helmholtz free energy is written as
\begin{equation}
F(V,T)
=
F_{\rm ig}(V,T)
+
F_{\rm el}(V,T)
+
F_{\rm xs}(V,T),
\end{equation}
where $F_{\rm ig}$ is the ideal atomic contribution, $F_{\rm el}$ is the
thermal electronic contribution, and $F_{\rm xs}$ is the excess interatomic
contribution.

The excess free energy is represented as a low-order expansion in Eulerian
finite strain and temperature,
\begin{equation}
F_{\rm xs}
=
\sum_i\sum_j
\frac{a_{ij}}{i!j!} f^i \theta^j
+
\sum_i a_{iL} f^i t ,
\end{equation}
where $f$ is the Eulerian finite strain, $\theta$ is the nonlinear thermal
variable, and $t$ is the linear-temperature variable. For the present fit,
$m=0.60$ is used in the definition of $\theta$.

The thermal electronic contributions are controlled by the electronic heat-capacity
amplitude, $\zeta$, and the volume-dependent electronic onset temperature,
\begin{equation}
\tau(V)
=
\tau_{\infty}
+
\left(\tau_0-\tau_{\infty}\right)
\left(\frac{V}{V_0}\right)^{-\eta}.
\end{equation}

All thermodynamic properties are computed from derivatives of the fitted
Helmholtz free-energy surface. For example, pressure is obtained from
\begin{equation}
P
=
-\left(\frac{\partial F}{\partial V}\right)_T .
\end{equation}
Entropy is obtained from
\begin{equation}
S
=
-\left(\frac{\partial F}{\partial T}\right)_V .
\end{equation}
The product of thermal expansivity and isothermal bulk modulus is obtained
from the mixed derivative,
\begin{equation}
\alpha K_T
=
\left(\frac{\partial P}{\partial T}\right)_V
=
-\left(\frac{\partial^2 F}{\partial V \partial T}\right).
\end{equation}
The same Helmholtz free-energy surface is used to derive the remaining
EOS quantities through the appropriate derivatives and standard
thermodynamic identities. This ensures internal thermodynamic
consistency throughout the MgSiO$_3$--H lookup table. The fitted
parameters are listed in Table~\ref{tab:eos_params}.

\begin{table*}[t]
\centering
\caption{Parameters for the supercritical $\mathrm{MgSiO}_3\mathrm{H}_4$ equation of state.}
\label{tab:eos_params}
\footnotesize
\setlength{\tabcolsep}{6pt}
\renewcommand{\arraystretch}{1.08}
\begin{tabular*}{\textwidth}{@{\extracolsep{\fill}} l c r l @{}}
\toprule
Parameter & Symbol & Value & Unit \\
\midrule
Reference volume & $V_0$ & 71.21599 & cm$^3$ mol$^{-1}$ \\
Reference temperature & $T_0$ & 4000.0 & K \\
Thermal exponent & $m$ & 0.60 & -- \\
Atoms per formula unit & $N$ & 9.0 & -- \\
\addlinespace

\multicolumn{4}{@{}l}{\textit{Excess coefficients}} \\
Energy coefficient & $a_{00}$ & -1426.73962 & kJ mol$^{-1}$ \\
Pressure coefficient & $a_{10}$ & -929.67740 & kJ mol$^{-1}$ \\
Bulk modulus coefficient & $a_{20}$ & 10163.14329 & kJ mol$^{-1}$ \\
Higher-order strain coefficient & $a_{30}$ & 6139.31016 & kJ mol$^{-1}$ \\
Thermal energy coefficient & $a_{01}$ & 1978.38149 & kJ mol$^{-1}$ \\
Thermal pressure coefficient & $a_{11}$ & 1893.08545 & kJ mol$^{-1}$ \\
Thermal strain coefficient & $a_{21}$ & 814.32873 & kJ mol$^{-1}$ \\
\addlinespace

\multicolumn{4}{@{}l}{\textit{Linear-temperature coefficients}} \\
Linear energy term & $a_{0L}$ & 1404.50416 & kJ mol$^{-1}$ \\
Linear strain term & $a_{1L}$ & -507.86613 & kJ mol$^{-1}$ \\
\addlinespace

\multicolumn{4}{@{}l}{\textit{Electronic parameters}} \\
Electronic heat-capacity amplitude & $\zeta$ & 0.01667178 & kJ mol$^{-1}$ K$^{-1}$ \\
Onset temperature at $V_0$ & $\tau_0$ & 1229.01101 & K \\
Large-volume onset temperature & $\tau_{\infty}$ & 848.19186 & K \\
Volume-dependence exponent & $\eta$ & 1.04986 & -- \\
\bottomrule
\end{tabular*}
\end{table*}

\section*{B. Simulation Data}
\label{Appendix:Data}

The temperature, pressure, and density points from our simulations can be found in Table~\ref{tab:raw_data}.

\begin{table*}[t]
\centering
\caption{Time-averaged pressure and density for DFT-MD simulations of
$\mathrm{MgSiO}_3\mathrm{H}_4$ at 4000, 6000, and 8000 K. Pressure
uncertainties are given after the $\pm$ symbol.}
\label{tab:raw_data}
\footnotesize
\setlength{\tabcolsep}{6pt}
\renewcommand{\arraystretch}{1.08}

\begin{tabular*}{\textwidth}{@{\extracolsep{\fill}} ll ll ll @{}}
\toprule
\multicolumn{2}{c}{$T = 4000$ K} &
\multicolumn{2}{c}{$T = 6000$ K} &
\multicolumn{2}{c}{$T = 8000$ K} \\
\cmidrule(lr){1-2}
\cmidrule(lr){3-4}
\cmidrule(l){5-6}
$P$ (GPa) & $\rho$ (g cm$^{-3}$) &
$P$ (GPa) & $\rho$ (g cm$^{-3}$) &
$P$ (GPa) & $\rho$ (g cm$^{-3}$) \\
\midrule
$339.98 \pm 0.84$ & 5.461 & $580.32 \pm 0.45$ & 6.422 & $615.83 \pm 0.47$ & 6.422 \\
$211.29 \pm 0.32$ & 4.682 & $368.43 \pm 1.0$  & 5.461 & $395.74 \pm 0.36$ & 5.461 \\
$124.77 \pm 0.40$ & 4.044 & $235.78 \pm 0.35$ & 4.682 & $257.90 \pm 0.56$ & 4.682 \\
$77.50 \pm 0.43$  & 3.518 & $145.62 \pm 0.35$ & 4.044 & $163.31 \pm 0.20$ & 4.044 \\
$47.64 \pm 0.25$  & 3.078 & $93.59 \pm 0.26$  & 3.518 & $107.52 \pm 0.37$ & 3.518 \\
$29.83 \pm 0.22$  & 2.709 & $59.88 \pm 0.17$  & 3.078 & $71.50 \pm 0.39$  & 3.078 \\
$18.48 \pm 0.16$  & 2.397 & $38.99 \pm 0.16$  & 2.709 & $48.16 \pm 0.35$  & 2.709 \\
$9.02 \pm 0.13$   & 2.013 & $25.48 \pm 0.15$  & 2.397 & $32.89 \pm 0.14$  & 2.397 \\
$5.64 \pm 0.13$   & 1.801 & $13.60 \pm 0.11$  & 2.013 & $18.77 \pm 0.13$  & 2.013 \\
$4.90 \pm 0.13$   & 1.618 & $9.38 \pm 0.10$   & 1.801 & $13.32 \pm 0.094$ & 1.801 \\
                  &       & $6.40 \pm 0.12$   & 1.618 & $9.69 \pm 0.13$   & 1.618 \\
\bottomrule
\end{tabular*}
\end{table*}

\clearpage  

 \bibliographystyle{aasjournal}
\bibliography{spm_references}

\clearpage  

\end{document}